%% file: main.tex
\documentclass[aps,prl,10pt,twocolumn,superscriptaddress,amsmath,amssymb,floatfix,footinbib,raggedbottom]{revtex4-2}

\usepackage{silence}
\usepackage{mathrsfs}
\usepackage{mathtools}
\usepackage{bm}
\usepackage{amssymb}
\usepackage{graphicx}
\usepackage[dvipsnames]{xcolor}

\definecolor{brickred}{rgb}{0.8, 0.25, 0.33}
\newcommand\myshade{85}
\colorlet{mylinkcolor}{BrickRed}
\colorlet{mycitecolor}{NavyBlue}
\colorlet{myurlcolor}{Aquamarine}

\usepackage{hyperref}
\hypersetup{
 linkcolor = mylinkcolor!\myshade!black,
 citecolor = mycitecolor!\myshade!black,
 urlcolor  = myurlcolor!\myshade!black,
 colorlinks = true,
}

\newcommand{\tr}{\operatorname{tr}}
\newcommand{\dd}{\mathrm{d}}
\newcommand{\TF}{\mathrm{TF}}

\newcommand{\ee}{\mathbf{e}}
\newcommand{\eu}{\mathbf{e}_u}
\newcommand{\kv}{\mathbf{k}}

\begin{document}

\title{
Phonon Angular Momentum Pumping in Curved Membranes
}

\author{Pablo A. Morales}
\email{pablo\_morales@araya.org}
\affiliation{Research Division, Araya Inc., Tokyo 101-0025, Japan}
\affiliation{Centre for Complexity Science, Imperial College London, London SW7 2AZ, UK}

\begin{abstract}
Flexural vibrations dominate the low-energy spectrum of a flat crystalline membrane but do not couple to phonon Hall viscosity. We show that static curvature couples normal motion to in-plane strain, allowing an oscillatory flexural drive to generate elliptical in-plane motion and pump phonon angular momentum along the surface normal. A net angular momentum persists in randomly rippled membranes, with its magnitude and handedness controlled by the drive and ripple spectra. Transfer of the pumped angular momentum to the support produces a steady torque that reverses with the magnetic field and scales with the mean-square curvature. Using the measured bulk Hall viscosity of $\alpha$-RuCl$_3$ to set the scale, we estimate a torque of order $10^{-23}$--$10^{-22}\,\mathrm{N\,m}$ for a few-layer membrane, within reach of demonstrated torsional sensors. The same flexural-to-shear response provides a probe of phonon Hall viscosity in atomically thin crystals.
\end{abstract}

\maketitle

\textit{Introduction.}---
In crystals with broken time-reversal ($\mathcal T$) symmetry, acoustic phonons can acquire a nondissipative handed response. In fluids the corresponding transport coefficient is Hall, or odd, viscosity, the antisymmetric part of the viscosity tensor, whose stress is proportional to strain rate but produces no power~\cite{1ASZ1995,Avron1998,Read2009,HoyosSon2012,HughesLeighFradkin2011}. Active and nonconservative solids admit a related static response, odd elasticity~\cite{Scheibner2020,Fruchart2023}, recently extended to curved manifolds~\cite{ZhouCurvedOdd2025}.
In passive crystals the acoustic analogue is phonon Hall viscosity, the leading $\mathcal{T}$-odd term in the long-wavelength phonon action. It gives acoustic bands Berry curvature, mixes longitudinal and transverse polarizations, and contributes to acoustic Faraday rotation and phonon Hall heat transport~\cite{Strohm2005,Sheng2006,Kagan2008,Qin2012,Savary2021,BarkeshliChungQi2011}.

Recent experiments have brought this coefficient into the acoustic regime. Ultrasonic measurements of the acoustic Faraday effect in the Kitaev magnet $\alpha$-RuCl$_3$ have determined its phonon Hall viscosity and linked it to the intrinsic thermal Hall response~\cite{Shragai2026}. The experiment probes sound propagation through a bulk crystal, leaving open how to measure the Hall response of an atomically thin membrane. Phonons also carry angular momentum that is odd in magnetization, as expected from the Einstein--de Haas effect~\cite{ZhangNiu2014,ZhangNiu2015,Garanin2015}, and this angular momentum has now been directly measured~\cite{ZhangNatPhys2025}.

A crystal suspended as a membrane adds a scalar flexural mode to this acoustic sector. A single flexural displacement defines no polarization plane, carries no angular momentum about the surface normal, and on a flat sheet decouples from the Hall action at Gaussian order. The dominant vibration of a flat membrane is therefore excluded from its Hall-active acoustic sector. On a curved surface, however, a normal displacement changes the in-plane metric at linear order and enters the Green strain on the same footing as an in-plane phonon~\cite{NelsonPiranWeinberg,LandauElasticity,Amorim2016}. 
The same geometric coupling hybridizes flexural and in-plane modes in ordinary elasticity~\cite{MoralesCastro2026}.

In this Letter we formulate the Hall-viscous dynamics of a curved
crystalline membrane. The Hall tensor projects the curvature-induced strain onto shear, producing a direct flexural vertex proportional to the tracefree extrinsic curvature, while mean curvature enters only through the longitudinal deformation generated by ordinary elasticity. Solving the in-plane dynamics driven by a prescribed normal displacement, we show that the drive generates a field-odd shear response and pumps in-plane phonon angular momentum. Transfer of the pumped angular momentum to the support produces a steady torque. The drive and ripple spectra control its handedness, which follows the Hall coefficient $\vartheta_H$ when the ripple-scattered waves are driven below acoustic resonance, and direct solutions for finite clamped membranes confirm its sign and magnitude. The linear cross-response provides a direct determination of $\vartheta_H$ in a suspended crystal.

\textit{Phonon Hall viscosity.}---In a flat crystal, phonon Hall viscosity contributes a term bilinear in the linear strain $U_{ij}=(\partial_i u_j+\partial_j u_i)/2$ and its time derivative~\cite{BarkeshliChungQi2011,Savary2021},
\begin{equation}
S_H^{\rm flat}=\frac12\int\dd t\dd^2x\,\vartheta_H^{ijkl}U_{ij}\partial_tU_{kl}.
\label{eq:SH_flat}
\end{equation}
Here $u_i$ is the lattice displacement and $\vartheta_H^{ijkl}$ is the Hall-viscosity tensor for symmetric strain, symmetric within each index pair and antisymmetric under exchange of the pairs. We retain its local isotropic shear component and extend it to a curved reference surface. The reference and deformed metrics, $g_{ab}$ and $g'_{ab}$, define the Green strain $E_{ab}=(g'_{ab}-g_{ab})/2$. Its part linear in displacement, $E_{ab}^{(1)}$, gives the quadratic Hall action
\begin{equation}
S_H^{(2)}=\frac{\vartheta_H}{2}\int\dd t\dd A\,\mathcal A^{abcd}E_{ab}^{(1)}\partial_tE_{cd}^{(1)},
\label{eq:SH_covariant}
\end{equation}
where the superscripts denote orders in displacement, $\dd A=\sqrt{\det g}\,\dd^2x$, and
\begin{equation}
\mathcal A^{abcd}=\frac{1}{4}\left(\epsilon^{ac}g^{bd}+\epsilon^{ad}g^{bc}+\epsilon^{bc}g^{ad}+\epsilon^{bd}g^{ac}\right).
\label{eq:hall_tensor}
\end{equation}
Surface indices are raised and lowered with $g_{ab}$, and $\epsilon^{ab}$ is the antisymmetric area tensor, with $\epsilon^{xy}=+1$ in an oriented orthonormal frame. The flat tensor in~\eqref{eq:SH_flat} is $\vartheta_H\mathcal A^{ijkl}$ evaluated at $g_{ij}=\delta_{ij}$. Since $\mathcal A^{abcd}g_{cd}=0$, this coupling acts only on shear. The coefficient $\vartheta_H$ has units of action per area and reverses with the microscopic $\mathcal T$-odd source.

With the force-balance convention $\rho\ddot u^a=\nabla_b(T_{\rm el}^{ab}+\sigma_H^{ab})$, where $\rho$ is the areal mass density and $T_{\rm el}^{ab}$ is the ordinary elastic stress, variation gives
\begin{equation}
\sigma_H^{ab}=-\vartheta_H\mathcal A^{abcd}\dot E_{cd}.
\label{eq:hall_stress}
\end{equation}
The Hall stress rotates the shear response without supplying energy, because antisymmetry under pair exchange makes its instantaneous power density vanish,
\begin{equation}
P_H=\sigma_H^{ab}\dot E_{ab}=-\vartheta_H\mathcal A^{abcd}\dot E_{ab}\dot E_{cd}=0.
\label{eq:zero_power}
\end{equation}
This is the passive response associated with strain-space Berry curvature, derived in End Matter, whereas a static odd elastic modulus describes the nonconservative response of Refs.~\cite{Scheibner2020,Fruchart2023,ZhouCurvedOdd2025}.

\begin{figure}[t]
\centering
\includegraphics[width=\columnwidth]{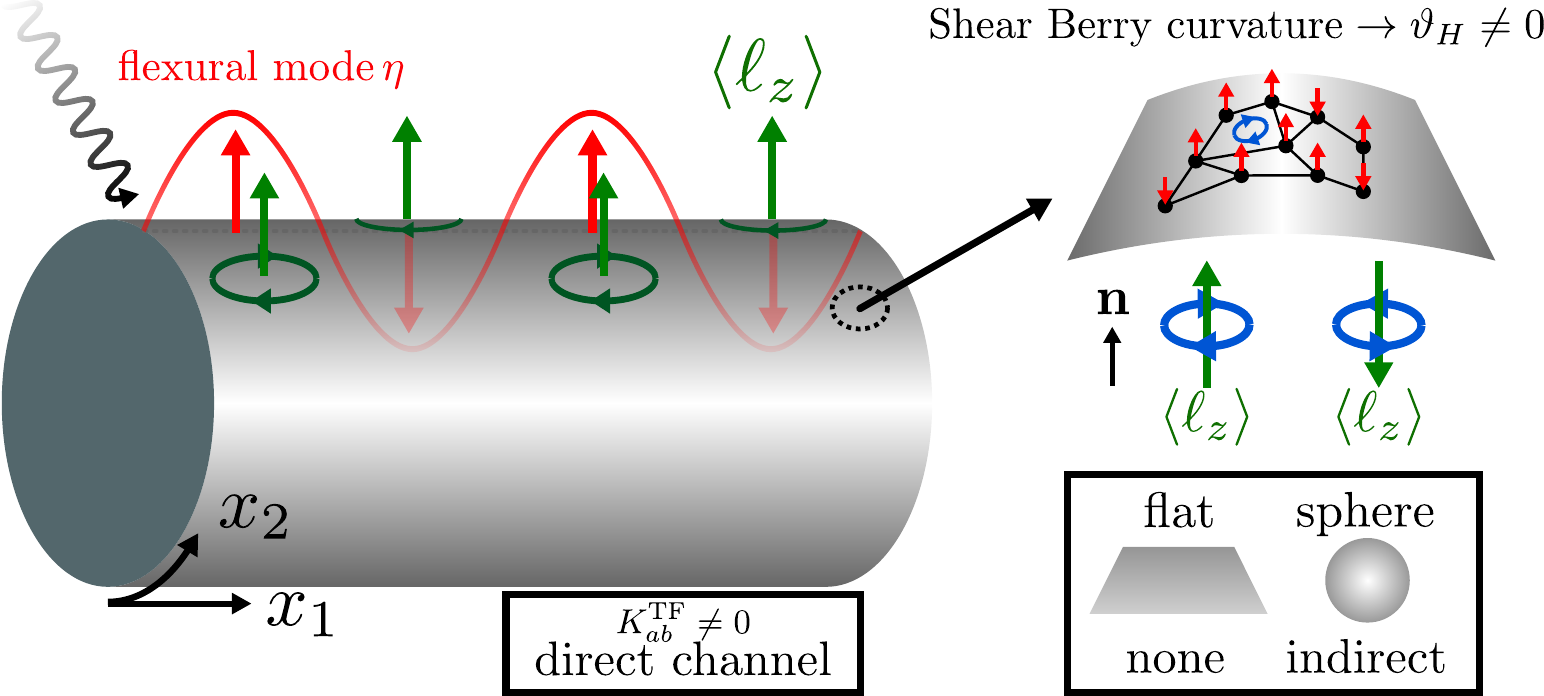}
\caption{\label{fig:rule}Curvature converts normal motion $\eta$ into Hall-active in-plane motion. A cylinder has a direct shear coupling, $K^{\TF}_{ab}\ne0$, while on a sphere, $K^{\TF}_{ab}=0$ and a nonuniform normal drive excites meridional shear that generates azimuthal Hall motion. The resulting ellipse carries local angular momentum along the normal, with handedness reversed by the $\mathcal T$-odd source.}
\end{figure}

\textit{Tracefree curvature transduction.}---For a reference surface $\Sigma$ embedded in $\mathbb R^3$, with tangents $\ee_a$ and unit normal $\mathbf n$, write the displaced position as $\mathbf X'(x)=\mathbf X(x)+u^a\ee_a+\eta\mathbf n$, where $u^a$ and $\eta$ are tangential and normal displacements. To first order in these fields, the strain is~\cite{NelsonPiranWeinberg,LandauElasticity,MoralesCastro2026}
\begin{equation}
E_{ab}^{(1)}=U_{ab}+\eta K_{ab},\qquad U_{ab}\equiv\nabla_{(a}u_{b)},
\label{eq:linear_strain}
\end{equation}
where $K_{ab}$ is the second fundamental form of $\Sigma$. The geometric strain $\eta K_{ab}$ produces the established hybridization of flexural and in-plane motion in ordinary elasticity. In the Hall action~\eqref{eq:SH_covariant}, its pure-flexural contribution vanishes because $\mathcal A^{abcd}K_{ab}K_{cd}=0$, while the mixed terms combine after integration by parts in time to give
\begin{equation}
S_{H,\rm dir}^{(2)}=\vartheta_H\!\int\!\dd t\dd A\,\eta\mathcal A^{abcd}K_{ab}^{\TF}\partial_tU_{cd},
\label{eq:index_transduction}
\end{equation}
with $K_{ab}^{\TF}=K_{ab}-\tfrac12Kg_{ab}$ and $K=g^{ab}K_{ab}$. Only directional differences in curvature enter this direct coupling, which vanishes on an umbilic surface that bends equally in every direction. Mean curvature instead generates longitudinal deformation whose shear couples to the in-plane Hall stress, as illustrated in Fig.~\ref{fig:rule}.

\textit{Coupled dynamics and phonon angular momentum.}---Combining the Hall coupling with the kinetic, stretching, and bending energies gives the quadratic Lagrangian density
\begin{equation}
\begin{aligned}\mathcal{L}={}&\frac{\rho}{2}\left(\dot u_a\dot u^a+\dot\eta^2\right)-\frac{\lambda}{2}(\tr E)^2-\mu E_{ab}E^{ab}\\ &-\frac{1}{2}\eta\mathcal L_{\rm HC}\eta+\vartheta_H\mathbf e\times\dot{\mathbf e}.\end{aligned}
\label{eq:full_lagrangian}
\end{equation}
In a local orthonormal frame, $\mathbf e=\eu+\eta\kv$ collects the two shear components, with $\eu=((U_{xx}-U_{yy})/2,U_{xy})$, $\kv=((K_{xx}-K_{yy})/2,K_{xy})$, and $\mathbf a\times\mathbf b=a_1b_2-a_2b_1$. The two-dimensional Lam\'e coefficients $\lambda,\mu$ determine the elastic stress $T_{\rm el}^{ab}=\lambda(\tr E)g^{ab}+2\mu E^{ab}$, where $\tr E=g^{ab}E_{ab}$. The normal operator $\mathcal L_{\rm HC}$ follows from the quadratic Helfrich--Canham bending energy~\cite{MoralesCastro2024}. It reduces to $\kappa\nabla^4$, with bending rigidity $\kappa$, in a flat tensionless membrane and includes curvature and tension terms on a curved reference surface. Varying~\eqref{eq:full_lagrangian} gives

\begin{equation}
\begin{aligned}\rho\ddot u^a &=\nabla_b T_{\rm el}^{ab}-\vartheta_H\mathcal{A}^{abcd}\nabla_b (\dot U_{cd}+\dot\eta K_{cd}), \\ \rho\ddot\eta &=-\mathcal L_{\rm HC}\eta-K_{ab}T_{\rm el}^{ab}+2\vartheta_H(\kv\times\dot{\ee}_u).\end{aligned}
\label{eq:motion}
\end{equation}
The term $-K_{ab}T_{\rm el}^{ab}$ couples normal motion to in-plane strain even at $\vartheta_H=0$, while the Hall force adds a transverse response that reverses with $\vartheta_H$.

The observable is the phonon angular momentum carried by the in-plane displacement about the local normal, taken as the $z$ direction~\cite{ZhangNiu2014,ZhangNiu2015,Garanin2015},
\begin{equation}
\ell_z=\rho (\mathbf{u}\times\dot{\mathbf{u}})_z=\rho (u_x\dot u_y-u_y\dot u_x).
\label{eq:Lz_density}
\end{equation}
For a mode $\mathbf u(t)=\operatorname{Re}[\mathbf a e^{-i\omega t}]$, angle brackets denote a period average, giving $\langle\ell_z\rangle=\rho\omega\operatorname{Im}(a_x^*a_y)$. This vanishes for linear polarization and equals $\pm\rho\omega|a_x|^2$ for circular polarization. In the flat bulk the in-plane Hall term $\vartheta_H \eu\times\dot{\ee}_u$ couples $u_x$ and $u_y$. For a plane wave $\mathbf u\propto e^{i(\mathbf q\cdot\mathbf r-\omega t)}$ it adds an antisymmetric imaginary piece $\propto i\omega\vartheta_H q^2$ to the dynamical matrix, mixing longitudinal and transverse motion with a relative phase and
rendering the eigenmodes elliptical. For these flat-crystal modes, the ellipticity and angular momentum per phonon scale with the dimensionless Hall mixing
\begin{equation}
\epsilon_H(\omega)=\frac{\vartheta_H \omega}{\mu},\qquad\frac{L_z^{\rm ph}}{\hbar}\sim\epsilon_H,
\label{eq:epsilon_H}
\end{equation}
where $L_z^{\rm ph}$ is the angular momentum per phonon. This is the established Hall-induced ellipticity of in-plane phonons~\cite{BarkeshliChungQi2011}, the membrane counterpart of the acoustic Faraday effect measured in bulk $\alpha$-RuCl$_3$~\cite{Shragai2026}. In a flat membrane, a purely normal drive cannot excite this in-plane sector within linear theory.

To determine how curvature enables this excitation, consider a normal drive $\eta\propto e^{i(qx^1-\omega t)}$ along a principal direction, with $K_{ab}=\operatorname{diag}(\mathfrak K_1,\mathfrak K_2)$, $\Delta K=\mathfrak K_1-\mathfrak K_2$, and $K_G=\mathfrak K_1\mathfrak K_2$. The leading local inverse propagators are
\begin{equation}
\begin{split}\mathcal D_L&=(\lambda+2\mu)q^2-\rho\omega^2-i\rho\Gamma_L\omega,\\\mathcal D_T&=\mu q^2-\rho\omega^2-i\rho\Gamma_T\omega,\end{split}
\label{eq:DLDT}
\end{equation}
where $\Gamma_L,\Gamma_T\geq0$ are the longitudinal and transverse damping rates. This approximation requires curvature varying slowly over a wavelength and $q^2\gg|K_G|$, as illustrated by the exact spherical spectrum in End Matter.

Let $u_L^{{}_{(0)}}$ and $u_T^H$ denote the complex longitudinal and transverse displacement amplitudes at zeroth and first order in $\vartheta_H$. Projecting~\eqref{eq:motion} along and perpendicular to the drive gives
\begin{equation}
\begin{aligned}\mathcal D_Lu_L^{{}_{(0)}}&=iq(\lambda K+2\mu\mathfrak K_1)\eta,\\\mathcal D_Tu_T^H&=\frac{\vartheta_H\omega q}{2}\left(\Delta K\eta+iqu_L^{{}_{(0)}}\right).\end{aligned}
\label{eq:driven_amplitudes}
\end{equation}
In the first, the geometric strain $\eta K_{ab}$ drives $u_L^{{}_{(0)}}$ through the gradient of its longitudinal stress $P\eta$, with $P=\lambda K+2\mu\mathfrak K_1$. In the second, the Hall stress $\sigma_H^{xy}=\vartheta_H\dot e_1$ converts the field-even shear $e_1^{{}_{(0)}}=(\Delta K\eta+iqu_L^{{}_{(0)}})/2$ into transverse motion. The two terms of $e_1^{{}_{(0)}}$ are the direct and indirect channels of Fig.~\ref{fig:rule}. Eliminating $u_L^{{}_{(0)}}$ and using $E_{xy}^H=iqu_T^H/2$ gives the field-odd shear susceptibility $\chi^H_{E_{xy},\eta}=E_{xy}^H/\eta$, derived in Supplementary Material (SM)~\cite{supp}, \S II,
\begin{equation}
\chi^H_{E_{xy},\eta}=\frac{i\vartheta_H\omega q^2}{4\mathcal D_T}\left[\Delta K-\frac{q^2P}{\mathcal D_L}\right].
\label{eq:cross_response}
\end{equation}
In the undamped elastic regime $\rho\omega^2\ll\mu q^2$, \eqref{eq:cross_response} reduces to $E_{xy}^H\simeq(i\epsilon_H/2)e_1^{{}_{(0)}}$. The Hall mixing that renders flat-crystal phonons elliptical thus sets the amplitude and phase of $E_{xy}^H$ relative to $e_1^{{}_{(0)}}$. Measuring the Hall shear against the imposed normal displacement determines $\vartheta_H$ for known geometry and elastic constants. 

With damping neglected, $\Gamma_L=\Gamma_T=0$, and away from acoustic resonances, the denominators in~\eqref{eq:DLDT} are real. The Hall-induced transverse motion then leads or lags the longitudinal motion by a quarter period. From $\langle\ell_z\rangle=\rho\omega\operatorname{Im}[(u_L^{{}_{(0)}})^*u_T^H]$, combining the longitudinal amplitude in~\eqref{eq:driven_amplitudes} with the Hall shear~\eqref{eq:cross_response} gives our main result for the local time-averaged angular momentum density pumped by the normal drive,
\begin{equation}
\begin{aligned}\langle\ell_z\rangle&=-2\rho\omega|\eta|^2\frac{P}{\mathcal D_L}\operatorname{Im}\chi^H_{E_{xy},\eta}\\&=-\frac{\rho\vartheta_H\omega^2q^2|\eta|^2}{2\mathcal D_L\mathcal D_T}P\left[\Delta K-\frac{q^2P}{\mathcal D_L}\right].\end{aligned}
\label{eq:Lz_general}
\end{equation}
The observable~\eqref{eq:Lz_density} gives a finite time average from linear displacement dynamics. Both the longitudinal amplitude and the Hall shear are linear in curvature, making~\eqref{eq:Lz_general} quadratic in $K_{ab}$ and zero in the flat limit. The angular momentum is invariant under $K_{ab}\to-K_{ab}$ at fixed orientation, whereas its handedness reverses with $\vartheta_H$. 

\textit{Geometry and net angular momentum.}---The indirect channel can be isolated on a sphere, where equal principal curvatures eliminate the direct Hall vertex. A quadrupolar normal drive still generates meridional shear and Hall-induced azimuthal motion, demonstrating local pumping without anisotropic curvature. On a complete sphere, the local angular momentum points along the normal and cancels between antipodal points, so a finite local response need not yield net angular momentum. 

In a randomly rippled membrane, the linear shear can average to zero while the quadratic angular momentum remains finite. To establish whether local contributions reinforce one another, consider statistically homogeneous, isotropic shallow ripples much longer than the drive wavelength. Write their curvature variance as $\sigma_K^2=\langle K_{ab}K^{ab}\rangle$ and the combined time and ripple average with double brackets. At leading order in $\rho\omega^2/\mu q^2$, \eqref{eq:Lz_general} holds for arbitrary drive direction with $\mathfrak K_1$ and $\mathfrak K_2$ replaced by the normal curvatures along and across it (SM~\cite{supp}, \S~III), averaging over the ripples at zero damping gives 
\begin{equation}
\langle\!\langle\ell_z\rangle\!\rangle=\frac{\rho\vartheta_H\omega^2q^2|\eta|^2\sigma_K^2}{4\mathcal D_L\mathcal D_T}\frac{(\lambda+\mu)(2\lambda+\mu)}{\lambda+2\mu}.
\label{eq:lz_disorder}
\end{equation}
Local contributions thus add, so a membrane that is flat on average carries a net angular momentum proportional to its mean-square curvature. The full inertial average retains the sign of $\vartheta_H$ below transverse resonance when $2\lambda+\mu>0$. Individual saddle patches can contribute with the opposite sign for $\lambda\geq0$, but the cross terms responsible cancel under homogeneous ripple averaging.

\begin{figure}[tbp]
\centering
\includegraphics[width=\columnwidth]{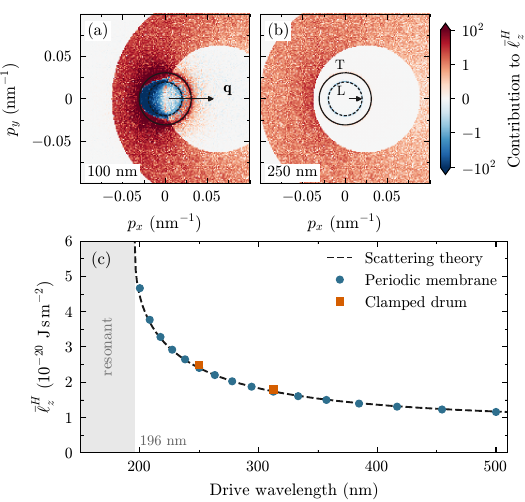}
\caption{\label{fig:numerics}
Ripple scattering at $10\,\mathrm{GHz}$. (a),(b) Contributions of tangential wavevectors $\mathbf p=\mathbf q+\mathbf Q$ for $100$ and $250\,\mathrm{nm}$ drives on one full-geometry realization, normalized to the mean per occupied wavevector in (b). The ripple band forms an annulus centered on $\mathbf q$, and circles mark longitudinal (L) and transverse (T) acoustic resonances. Primary scattering outside T carries the Hall sign, while higher-order scattering on the resonance circles in (b) contributes $-1\%$. (c) Field-odd angular momentum per area $\bar\ell_z^H$, scaled to a mean-square normal displacement of $1\,\mathrm{nm}^2$, from scattering theory and averages of four periodic realizations and of three clamped drums with a $1\,\mu\mathrm m$ drive taper. Standard deviations lie within the symbols, and shading marks primary scattering onto T below $196\,\mathrm{nm}$.}
\end{figure}

When the ripple and drive wavelengths become comparable, the response must resolve scattering by the curvature. A ripple wavevector $\mathbf Q$ scatters the normal drive $\mathbf q$ into a tangential wave $\mathbf p=\mathbf q+\mathbf Q$. Solving the driven equations to leading order in ripple height and Hall viscosity, while retaining arbitrary $Q/q$, gives the finite-damping condition

\begin{equation}
c_T^2|\mathbf q+\mathbf Q|^2>\omega^2+\Gamma_L\Gamma_T,\qquad \lambda\geq0,
\label{eq:scattering_condition}
\end{equation}
where $c_T=\sqrt{\mu/\rho}$ is the transverse sound speed. When this inequality holds throughout the ripple spectrum, all nonzero contributions to the averaged angular momentum have the sign of $\vartheta_H$, as derived in SM~\cite{supp}, \S V. In the weak-damping limit, it places the drive below the transverse resonance $\omega=c_T|\mathbf p|$. Crossing that resonance changes the phase of the transverse response relative to the curvature force and permits a reversal of the angular momentum. Distinct ripple components add in a periodic spatial average or a statistically homogeneous ensemble, so the sign condition requires neither isotropy nor Gaussian disorder and recovers the local result for $Q\ll q$. A finite drive envelope mixes incident wavevectors and allows interference, motivating the clamped-membrane calculation below.

To test this spectral criterion beyond the local approximation, we solve the full covariant tangential equations on periodic sheets and clamped square membranes. Figure~\ref{fig:numerics} uses $5\,\mu\mathrm m$ membranes with $50$--$100\,\mathrm{nm}$ ripples, a $10\,\mathrm{GHz}$ normal drive, and $\Gamma_L=\Gamma_T=10^{-3}\omega$. A $100\,\mathrm{nm}$ drive scatters across both acoustic resonance circles, producing nearly cancelling contributions and sample-dependent signs. Lengthening the drive to $250\,\mathrm{nm}$ satisfies~\eqref{eq:scattering_condition}. Parameters and spectral and finite-element methods are given in SM~\cite{supp}, \S V.

The periodic solutions agree quantitatively with scattering theory from $208$ to $500\,\mathrm{nm}$. At $250\,\mathrm{nm}$, the clamped response has the Hall sign and, after drive normalization, exceeds theory by a few percent, mainly due to weakly excited drum resonances. The agreement establishes a net response in finite clamped membranes despite the local interference allowed by the drive envelope.

\textit{Microscopic origin, magnitude, and experiment.}---The coefficient $\vartheta_H$ is the shear Berry curvature of the gapped microscopic sector. Its Kubo formula~\eqref{eq:kubo} involves the stress operators $T_\alpha=\partial H_{\rm micro}/\partial e_\alpha$, where $H_{\rm micro}$ is the microscopic Hamiltonian and $e_\alpha$, $\alpha=1,2$, are the shear components~\cite{Savary2021,ZhangSpinon2021}. Broken time reversal and a magnetic point group allowing the in-plane Hall coefficient are required~\cite{Savary2021}.
A candidate is the layered magnet $\alpha$-RuCl$_3$, whose bond-dependent exchange~\cite{Winter2017,Takagi2019} couples strain to these stress operators~\cite{Dhakal2024}. The measured bulk value $\vartheta_H^{\rm 3D}\simeq1.3\times10^{-5}\,\mathrm{Pa\,s}$~\cite{Shragai2026} gives the five-layer scale $\vartheta_H\simeq5d\vartheta_H^{\rm 3D}=4\times10^{-14}\,\mathrm{kg\,s^{-1}}$ at layer thickness $d=0.6\,\mathrm{nm}$, used in Fig.~\ref{fig:numerics}. The measured component mixes in-plane and out-of-plane polarizations, and its few-layer persistence is unknown, so this sets a scale for the membrane coefficient. Adiabaticity requires $\hbar\omega$ and $k_BT$ below the gap of stress-coupled excitations. At $10\,\mathrm{GHz}$, $\hbar\omega\simeq0.04\,\mathrm{meV}$ lies below the meV magnetic scale~\cite{Winter2017,Takagi2019}, suggesting temperatures well below $10\,\mathrm K$ and fields away from gap closure. Lowering frequency improves this separation but reduces the shear linearly and angular momentum quadratically at fixed drive amplitude away from resonance.

Suspended magnetic membranes already allow optical actuation and interferometric detection of normal motion~\cite{Siskins2020}. A cylindrical membrane converts this motion into a calibrated Hall shear. For radius $R$ and a drive along the axis, $\mathfrak K_1=0$, $\mathfrak K_2=1/R$, and the intrinsic metric is flat, making~\eqref{eq:cross_response} exact for the tangential response. In the off-resonant elastic regime $\rho\omega^2\ll\mu q^2$, it reduces to
\begin{equation}
\frac{|U^H_{xy}|}{|\eta|/R}=\frac{|\vartheta_H|\omega}{2\mu}\frac{\lambda+\mu}{\lambda+2\mu}.
\label{eq:viscometer}
\end{equation}
Here $U_{xy}^H=E_{xy}^H$ because the curvature tensor is diagonal. With the geometric strain $|\eta|/R$, elastic constants, and frequency measured independently,~\eqref{eq:viscometer} determines $\vartheta_H$ in an atomically thin membrane (see SM~\cite{supp}, \S II, for the cylinder solution). For $R\sim1\,\mu\mathrm m$ and $|\eta|\sim1\,\mathrm{nm}$, the shear is of order $10^{-8}$ at the bulk-based Hall scale.

A second observable is the steady torque on the support. By~\eqref{eq:zero_power}, the Hall stress does no work, and the symmetric stresses exert no internal torque, so the drive supplies the dissipated energy and, with the clamp, balances the angular-momentum loss. A support-damping force $-\rho\Gamma_{\rm supp}\dot{\mathbf u}$ acts on material displaced by $\mathbf u+\eta\mathbf n$. If the support receives the removed angular momentum, its torque about the laboratory axis $\hat{\mathbf z}$ is
\begin{equation}
\tau_z=\Gamma_{\rm supp}\int\dd A\,\rho\langle(\mathbf u+\eta\mathbf n)\times\dot{\mathbf u}\rangle\cdot\hat{\mathbf z}.
\label{eq:torque}
\end{equation}
The normal displacement changes the lever arm at the same ripple order as the tangential contribution. The transfer model and lever-arm correction are derived in SM~\cite{supp}, \S III and \S V, where they are distinguished from a boundary-traction calculation. Other damping channels receive their corresponding fractions, with magnetic damping transferring angular momentum to the spin sector. Mechanical quality factors determine the total rate $\Gamma_{\rm tot}$, while phonon linewidths, ultrasound attenuation, and spin-lattice calculations~\cite{Dhakal2024} constrain magnetic losses. Their field dependence, combined with the dependence on size and clamping, can separate the support fraction $\Gamma_{\rm supp}/\Gamma_{\rm tot}$.

For the five-layer parameters of Fig.~\ref{fig:numerics}, the clamped mean density at $250\,\mathrm{nm}$ and $\eta_0=1\,\mathrm{nm}$ is $1.43\times10^{-20}\,\mathrm{J\,s\,m^{-2}}$. If support damping dominates,~\eqref{eq:torque} gives $1.4\times10^{-23}\,\mathrm{N\,m}$, including the calculated lever arm. Demonstrated torsional sensors reach $3\times10^{-24}\,\mathrm{N\,m\,Hz^{-1/2}}$ at millikelvin temperatures~\cite{Kim2016torque}. The estimate exceeds this noise-equivalent torque in a one-hertz bandwidth before membrane-to-sensor coupling losses, which remain device dependent.
Antisymmetrizing the torque or shear under $\mathbf B\to-\mathbf B$~\cite{Shragai2026} removes the field-even driven background. Equilibrium phonon angular momentum is also field odd~\cite{ZhangNiu2014,ZhangNatPhys2025}, but the pumped torque scales quadratically with drive amplitude and ripple height and vanishes on a flat drum within linear theory. Curvature reversal preserves the torque and reverses the shear, while varying the drive wavelength tests the resonance dependence in~\eqref{eq:scattering_condition}.

A normal drive on a curved membrane produces a linear Hall shear and a quadratic angular momentum, which survives random corrugation and can be transferred to the support as a steady torque. Both observables arise in suspended crystals with a symmetry-allowed phonon Hall viscosity, from magnetic insulators to two-dimensional electronic systems in a field~\cite{BarkeshliChungQi2011}. On a flat sheet, the dominant vibration of a membrane decouples from phonon Hall viscosity. Curvature turns it into a probe of that viscosity and a source of phonon angular momentum

\textit{Acknowledgments.}---I thank Piotr Surówka and Andrey Sadofyev for useful comments and discussions.

\bibliography{references}
\bibliographystyle{apsrev4-2}

\clearpage
\section*{End Matter}

\textit{Berry origin and shear representation.}---In an oriented orthonormal frame, strain separates into dilatation and shear,
\begin{equation}
e_0=\tfrac12(E_{xx}+E_{yy}),\quad e_1=\tfrac12(E_{xx}-E_{yy}),\quad e_2=E_{xy}.
\label{eq:e_components}
\end{equation}
Under a frame rotation by $\theta$, $e_1+ie_2$ rotates by $2\theta$. The rotationally invariant, orientation-odd Berry term is therefore the area form in this shear plane. A gapped microscopic sector following strain adiabatically contributes $S_B=\hbar\int\dd t\,i\langle\Psi_0(E)|\partial_t\Psi_0(E)\rangle$. Expanding its Berry connection to first order in strain gives
\begin{equation}
S_H=\vartheta_H\int\dd t\dd A\,(e_1\dot e_2-e_2\dot e_1)=2\vartheta_H\int\dd t\dd A\,|e|^2\dot\varphi,
\label{eq:SH_shear}
\end{equation}
where $e_1+ie_2=|e|e^{2i\varphi}$. The tensor representation follows from the rotation $(\mathcal J S)_{ab}=\tfrac12(\epsilon_a{}^cS_{cb}+\epsilon_b{}^cS_{ca})$ on tracefree symmetric tensors. It satisfies $\mathcal J^2=-1$ and $(\mathcal J S)^{ab}=\mathcal A^{abcd}S_{cd}$, giving~\eqref{eq:hall_tensor}. Thus the Hall stress rotates shear rates without coupling to dilatation. For a membrane of area $A_\Sigma$, let $|0\rangle=|\Psi_0\rangle$ be the nondegenerate microscopic ground state and $|n\rangle$ its excited states, with energies $E_0$ and $E_n$. The Berry connection $a_\alpha=i\langle\Psi_0|\partial_{e_\alpha}\Psi_0\rangle$ defines the curvature $\Omega_{12}^{\rm tot}=\partial_{e_1}a_2-\partial_{e_2}a_1$. In terms of the stress operators $T_\alpha=\partial_{e_\alpha}H_{\rm micro}|_{e=0}$, the Hall coefficient is
\begin{equation}
\begin{split}\vartheta_H&=\frac{\hbar}{2A_\Sigma}\Omega_{12}^{\rm tot}\\&=-\frac{\hbar}{A_\Sigma}\operatorname{Im}\sum_{n\ne0}\frac{\langle0|T_1|n\rangle\langle n|T_2|0\rangle}{(E_n-E_0)^2}.\end{split}
\label{eq:kubo}
\end{equation}
Strong magnetoelastic coupling and soft stress-coupled excitations can enhance this coefficient~\cite{Savary2021,ZhangSpinon2021}, provided the drive remains adiabatic.

For the curved strain~\eqref{eq:linear_strain}, $\mathbf e=\eu+\eta\kv$. The mixed part of~\eqref{eq:SH_shear}, after integration by parts at fixed geometry, is $\mathcal L_{H,\rm dir}=2\vartheta_H\eta\kv\times\dot{\ee}_u$, equivalent to~\eqref{eq:index_transduction}. In the principal frame the normal displacement shears along the curvature axes, and the Hall vertex couples it to $\dot U_{12}$, which rotates those axes.

\textit{Intrinsic-curvature correction.}---The local response~\eqref{eq:cross_response} neglects changes in the in-plane modes at relative order $K_G/q^2$. A sphere permits an exact calculation of these changes because its longitudinal and transverse vector harmonics diagonalize the covariant elastic operator, giving
\begin{equation}
\begin{split}\mathcal D_L^{\rm sph}&=(\lambda+2\mu)q^2-2\mu K_G-\rho\omega^2-i\rho\Gamma_L\omega,\\\mathcal D_T^{\rm sph}&=\mu(q^2-2K_G)-\rho\omega^2-i\rho\Gamma_T\omega,\end{split}
\label{eq:EM_sphere}
\end{equation}
where $R$ is the sphere radius, $\ell$ is the spherical-harmonic degree, $q^2=\ell(\ell+1)/R^2$, and $K_G=R^{-2}$. The shift $-2\mu K_G$ contains both the Ricci term in the elastic operator and the vector-Laplacian correction. The mean-square shear per mean-square tangential displacement is $(q^2-2K_G)/4$, compared with $q^2/4$ for a flat plane wave. At $\ell=1$, the toroidal mode is a shear-free rigid rotation and the longitudinal mode has compression alone. Consequently the Hall coupling also contains $\vartheta_H\omega(q^2-2K_G)/2$. In the static limit this factor cancels the transverse softening, leaving only a finite longitudinal correction. The derivation and quadrupolar response are given in SM~\cite{supp}, \S II and \S IV.

\input{supp_mat}

\end{document}

%% file: supp_mat.tex
\onecolumngrid
\clearpage
\appendix

\begin{center}
{\large Phonon Angular Momentum Pumping in Curved Membranes}
\vspace{1em} \\
{\large \textbf{Supplementary Material}}
\vspace{1em} \\
Pablo A. Morales \\
    \textit{Research Division, Araya Inc., Tokyo, Japan \\
    Centre for Complexity Science, Imperial College London, London, UK}
\end{center}

\renewcommand\appendixname{Supplementary Material}
\setcounter{figure}{0}
\renewcommand{\thefigure}{S\arabic{figure}}
\renewcommand{\thesection}{\arabic{section}}
\renewcommand{\theequation}{S.\arabic{equation}}
\renewcommand{\thepage}{S\arabic{page}}
\renewcommand{\theHfigure}{S\arabic{figure}}
\setcounter{page}{1}

\setcounter{equation}{0}

Section~I fixes the symmetry allowance and scale of $\vartheta_H$, and Sec.~II derives the shear response~\eqref{eq:cross_response} with the exact cylinder and sphere solutions behind~\eqref{eq:viscometer} and Fig.~\ref{fig:rule}. Section~III derives the angular momentum density~\eqref{eq:Lz_general}, its ripple average~\eqref{eq:lz_disorder}, and the torque balance behind~\eqref{eq:torque}, Sec.~IV the spherical spectrum~\eqref{eq:EM_sphere}, and Sec.~V the sign condition~\eqref{eq:scattering_condition} and the numerical results of Fig.~\ref{fig:numerics}. Unprefixed equations refer to the main text or its End Matter.

\section{I. Passive Berry-curvature origin, symmetry allowance, and scale}
 
The microscopic definition and passivity of $\vartheta_H$ are given by~\eqref{eq:kubo} and~\eqref{eq:zero_power}. As a simple scale check, $H_{\rm cell}=-\tfrac12(\Delta\sigma_z+ge_1\sigma_x+ge_2\sigma_y)$ gives $|\vartheta_H|=\hbar g^2/(4A_{\rm cell}\Delta^2)$, where $\sigma_i$ are Pauli matrices, $\Delta$ is the level splitting at zero strain, $g$ is the shear coupling, and $A_{\rm cell}$ is the cell area, with the sign fixed by the shear-coupling orientation and $\operatorname{sgn}\Delta$. The bilinear $e_1\dot e_2-e_2\dot e_1$ is odd under time reversal and orientation-reversing in-plane symmetries. The magnetic point group must allow its coefficient; broken time reversal alone does not suffice. An out-of-plane field supplies the axial quantity in the isotropic theory. In a honeycomb magnet, strain derivatives of bond-dependent exchange give the stress doublet $T_1,T_2$ in~\eqref{eq:kubo}; spin-lattice calculations determine these couplings and their symmetry allowance~\cite{Savary2021,Dhakal2024}.

The bulk acoustic-Faraday value $\vartheta_H^{\rm 3D}\simeq1.3\times10^{-5}\,\mathrm{Pa\,s}$~\cite{Shragai2026} gives $\vartheta_H\simeq d\vartheta_H^{\rm 3D}\simeq8\times10^{-15}\,\mathrm{kg\,s^{-1}}$ at $d=0.6\,\mathrm{nm}$, or $\vartheta_HA_{\rm cell}/\hbar\sim20$--$30$ for $A_{\rm cell}=0.3\,\mathrm{nm}^2$. This conversion assumes both comparable tensor components and persistence to few-layer thickness: the measured bulk component mixes in-plane and out-of-plane polarizations, whereas the membrane uses in-plane shear. At $10\,\mathrm{GHz}$ the resulting $|\vartheta_H|\omega/\mu\lesssim10^{-4}$ justifies the first-order Hall expansion, while the main text discusses the adiabatic gap and experimental limitations.

\section{II. Shear response and exact cylinder and sphere solutions}
 
We use the strain~\eqref{eq:linear_strain} and elastic energy density $\tfrac12\lambda(\operatorname{tr}E)^2+\mu E_{ab}E^{ab}$, with $c_L^2=(\lambda+2\mu)/\rho$ and $c_T^2=\mu/\rho$. In a principal frame $K_{12}=0$, the curvatures and their trace and difference are those of the main text.

Consider a mode propagating along $x_1$, with $u_a(x_1,t)$ and an imposed flexural oscillation $\eta(x_1,t)\propto e^{i(qx_1-\omega t)}$. The linearized Green strain $E_{ab}=\nabla_{(a}u_{b)}+\eta K_{ab}$ has components
\begin{equation}
E_{11}=iqu_1+\mathfrak{K}_1\eta,\qquad E_{22}=\mathfrak{K}_2\eta,\qquad E_{12}=\frac{i}{2}qu_2.
\label{eq:strain_components}
\end{equation}
The local patch uses the shear components~\eqref{eq:e_components}, $e_1=\tfrac12(iqu_1+\Delta K\eta)$ and $e_2=\tfrac12iqu_2$. Varying the quadratic in-plane elastic action gives the covariant phonon operator
\begin{equation}
\mathscr D^{a}{}_{b}=-\mu\Delta\delta^{a}{}_{b}-(\lambda+\mu)\nabla^{a}\nabla_{b}-\mu R^{a}{}_{b},
\label{eq:covariant_phonon_operator}
\end{equation}
where $\Delta=\nabla^c\nabla_c$ is the covariant Laplacian acting on tangent vectors and $R^a{}_b$ is the Ricci tensor, with the static Green function derived in Ref.~\cite{MoralesCastro2026}. Replacing the covariant derivatives at the origin of Riemann normal coordinates by plane-wave derivatives, and using $R^{a}{}_{b}=K_G\delta^{a}{}_{b}$, gives the frozen symbol
\begin{equation}
\mathscr D^{a}{}_{b}(q)=\mu(q^2-K_G)\delta^{a}{}_{b}+(\lambda+\mu)q^{a}q_{b}.
\label{eq:local_phonon_symbol}
\end{equation}
Connection and vector-Laplacian terms contribute at the same relative order $K_G/q^2$ as the explicit curvature term, so the frozen symbol alone cannot determine the spectrum at that order. We therefore retain only the leading denominators~\eqref{eq:DLDT}, requiring slowly varying curvature and $q^2\gg|K_G|$, which are exact on the cylinder and corrected on the sphere in Sec.~IV.

At $\vartheta_H=0$, the longitudinal response relates the in-plane stretch to the curvature drive,
\begin{equation}
u_1^{(0)}=\frac{iq(\lambda K+2\mu \mathfrak{K}_1)}{\mathcal D_L}\eta.
\label{eq:u1_even}
\end{equation}
The transverse response follows from the Hall force $\partial_1\sigma_H^{12}=iq\sigma_H^{12}$, with $\sigma_H^{12}=-i\omega\vartheta_H e_1$ evaluated on~\eqref{eq:u1_even}. To first order in $\vartheta_H$,
\begin{equation}
u_2^H=\frac{\vartheta_H\omega q}{2\mathcal D_T}\left[\Delta K-\frac{q^2(\lambda K+2\mu \mathfrak{K}_1)}{\mathcal D_L}\right]\eta.
\label{eq:u2_general}
\end{equation}
Multiplying~\eqref{eq:u2_general} by $iq/2$ gives the shear response~\eqref{eq:cross_response}.

Using $2\mathfrak K_1=K+\Delta K$, the bracket in~\eqref{eq:cross_response} can be written as $\Delta K(1-\mu q^2/\mathcal D_L)-(\lambda+\mu)Kq^2/\mathcal D_L$.
The first term contains the direct contribution and the tracefree part of the longitudinal elastic response, while mean curvature enters only through the second term, which alone survives on an umbilic patch, $\mathfrak K_1=\mathfrak K_2$. The exact spherical solution below establishes this umbilic response beyond the local approximation.

\paragraph{Axisymmetric quadrupole on a sphere.}
The shell-mediated response can be verified directly in spherical coordinates. Consider a sphere of radius $R$ with $K_{ab}=g_{ab}/R$ and the axisymmetric quadrupolar deformation
\begin{equation}
\eta(\theta,t)=\eta_2(t)P_2(\cos\theta),\qquad u_{\hat\theta}(\theta,t)=U_2(t)\partial_\theta P_2(\cos\theta),\qquad u_{\hat\phi}(\theta,t)=V_2(t)\partial_\theta P_2(\cos\theta).
\label{eq:sphere_ansatz}
\end{equation}
The hatted indices denote the local orthonormal basis. With $P_2(\cos\theta)=(3\cos^2\theta-1)/2$ and
\begin{equation}
g_2(\theta)\equiv\partial_\theta^2P_2-\cot\theta\,\partial_\theta P_2=3\sin^2\theta,
\label{eq:sphere_g2}
\end{equation}
the trace and shear components are
\begin{equation}
\operatorname{tr}E=\frac{2\eta_2-6U_2}{R}P_2,\qquad e_1=\frac{U_2}{2R}g_2,\qquad e_2=\frac{V_2}{2R}g_2.
\label{eq:sphere_strain}
\end{equation}
Although the normal displacement enters only the trace in~\eqref{eq:sphere_strain}, its spatial variation drives the meridional amplitude $U_2$, whose strain contains the shear $e_1$.

Using $\int\dd\Omega\,P_2^2=4\pi/5$, $\int\dd\Omega\,(\partial_\theta P_2)^2=24\pi/5$, and $\int\dd\Omega\,g_2^2=96\pi/5$, the tangential part of the reduced Lagrangian, divided by $4\pi/5$, is
\begin{equation}
\mathcal L_{2,\parallel}=3\rho R^2(\dot U_2^2+\dot V_2^2)-\frac{\lambda+\mu}{2}(2\eta_2-6U_2)^2-12\mu(U_2^2+V_2^2)+6\vartheta_H(U_2\dot V_2-V_2\dot U_2).
\end{equation}
For harmonic motion proportional to $e^{-i\omega t}$, the amplitudes obey
\begin{equation}
\mathcal D_{L,2}^{\rm sph}U_2+2i\omega\vartheta_H V_2=2(\lambda+\mu)\eta_2,\qquad \mathcal D_{T,2}^{\rm sph}V_2-2i\omega\vartheta_H U_2=0,
\end{equation}
where
\begin{equation}
\mathcal D_{L,2}^{\rm sph}=6\lambda+10\mu-\rho R^2\omega^2-i\rho R^2\Gamma_L\omega,\qquad \mathcal D_{T,2}^{\rm sph}=4\mu-\rho R^2\omega^2-i\rho R^2\Gamma_T\omega.
\label{eq:sphere_denominators}
\end{equation}
These are $R^2$ times the exact inverse propagators of the spherical $\ell=2$ longitudinal and transverse vector harmonics. Relative to flat denominators at $q^2=6/R^2$, each contains a shift $-2\mu/R^2$, derived for all $\ell$ in Sec.~IV.
To first order in $\vartheta_H$,
\begin{equation}
U_2^{(0)}=\frac{2(\lambda+\mu)}{\mathcal D_{L,2}^{\rm sph}}\eta_2,\qquad V_2^H=\frac{4i\omega\vartheta_H(\lambda+\mu)}{\mathcal D_{L,2}^{\rm sph}\mathcal D_{T,2}^{\rm sph}}\eta_2.
\end{equation}
The quadrupolar deformation therefore produces the Hall shear
\begin{equation}
E_{\hat\theta\hat\phi}^{H}(\theta)=\frac{6i\omega\vartheta_H(\lambda+\mu)}{R\mathcal D_{L,2}^{\rm sph}\mathcal D_{T,2}^{\rm sph}}\eta_2\sin^2\theta.
\label{eq:sphere_hall_shear}
\end{equation}
The direct geometric shear remains zero because $K_{ab}^{\rm TF}=0$. The response~\eqref{eq:sphere_hall_shear} is generated entirely by the meridional shell strain in~\eqref{eq:sphere_strain}. For zero damping, the local angular momentum density normal to the sphere is
\begin{equation}
\langle\ell_\perp(\theta)\rangle=\frac{8\rho\vartheta_H\omega^2(\lambda+\mu)^2}{(\mathcal D_{L,2}^{\rm sph})^2\mathcal D_{T,2}^{\rm sph}}|\eta_2|^2\left[\partial_\theta P_2(\cos\theta)\right]^2.
\end{equation}
On a complete sphere, $[\partial_\theta P_2]^2$ is invariant under the antipodal map while the outward normal changes sign, so the vector integral $\int\dd A\,\langle\ell_\perp\rangle\mathbf n$ vanishes. A spherical cap or an asymmetric drive does not enforce this cancellation.

\paragraph{Cylindrical geometry.}
For a cylinder, $\mathfrak{K}_1=0$, $\mathfrak{K}_2=1/R$, $K_G=0$, and $\Delta K=-1/R$, so~\eqref{eq:u2_general} and~\eqref{eq:cross_response} reduce to
\begin{equation}
u_2^H=-\frac{\vartheta_H\omega q}{2R\mathcal D_T}\left(1+\frac{\lambda q^2}{\mathcal D_L}\right)\eta,\qquad \chi^H_{E_{xy},\eta}=-\frac{i\vartheta_H\omega q^2}{4R\mathcal D_T}\left(1+\frac{\lambda q^2}{\mathcal D_L}\right).
\label{eq:cylinder}
\end{equation}
The intrinsic metric and curvature components are constant, so~\eqref{eq:cylinder} is exact for the tangential response. It vanishes for $R^{-1},\vartheta_H$, or $\omega\to0$ and reverses with either the field or curvature at fixed orientation. In the off-resonant elastic limit it gives the viscometer relation~\eqref{eq:viscometer}. At zero damping, the angular momentum of the cylinder follows from~\eqref{eq:u1_even},~\eqref{eq:cylinder}, and $\langle\ell_z\rangle=\rho\omega\operatorname{Im}(u_1^*u_2)$,
\begin{equation}
\langle\ell_z\rangle=\frac{\rho\lambda\vartheta_H\omega^2q^2|\eta|^2}{2R^2\mathcal D_L\mathcal D_T}\left(1+\frac{\lambda q^2}{\mathcal D_L}\right).
\label{eq:lz}
\end{equation}
For $\lambda>0$ below transverse resonance its sign follows $\vartheta_H$, and at $\lambda=0$ the axial drive has no longitudinal response and hence no angular momentum at this order. The two curvature insertions make angular momentum even under $K_{ab}\to-K_{ab}$ at fixed orientation, while the shear response is odd.

If an external normal force density $f_\eta$ is supplied in place of an imposed displacement, the flexural dynamics, which also contains the Helfrich--Canham stiffness, gives the normal response
\begin{equation}
\eta(q,\omega)=\frac{f_\eta(q,\omega)}{\mathcal E_{\rm HC}(q)+\mathcal E_{\rm st}(q,\omega)-\rho\omega^2-i\gamma_\eta\omega}.
\label{eq:eta_green}
\end{equation}
Here $\mathcal E_{\rm HC}$ is the Helfrich--Canham bending kernel, $\mathcal E_{\rm st}$ is the stretching kernel obtained by integrating the in-plane phonons out of the shell coupling $-K_{ab}T_{\rm el}^{ab}$, exactly as in the $\mathcal{T}$-even theory~\cite{MoralesCastro2026}, and $\gamma_\eta$ is the normal viscous-damping coefficient per area. This susceptibility fixes the amplitude for a given actuator, while the tangential calculation above takes $\eta$ as prescribed.

\section{III. Angular momentum density, ripple average, and steady-state balance}

The principal-axis solution~\eqref{eq:driven_amplitudes} generalizes to an arbitrary patch orientation. Let the mode propagate along $x_1$ and let the curvature tensor have components $K_{11},K_{22},K_{12}$ in this frame, with $K=K_{11}+K_{22}$ and $\Delta K=K_{11}-K_{22}$. The strain components are $E_{11}=iqu_1+K_{11}\eta$, $E_{22}=K_{22}\eta$, $E_{12}=\tfrac{i}{2}qu_2+K_{12}\eta$, so the Hall-active doublet is $e_1=\tfrac12(iqu_1+\Delta K\eta)$ and $e_2=\tfrac{i}{2}qu_2+K_{12}\eta$. At $\vartheta_H=0$ the driven amplitudes are
\begin{equation}
u_1^{(0)}=\frac{iqP}{\mathcal D_L}\eta,\qquad u_2^{(0)}=\frac{iqS}{\mathcal D_T}\eta,\qquad P=\lambda K+2\mu K_{11},\qquad S=2\mu K_{12},
\label{eq:S_zeroth_general}
\end{equation}
and the Hall forces $f_1^H=-q\omega\vartheta_H e_2^{(0)}$ and $f_2^H=q\omega\vartheta_H e_1^{(0)}$ add the first-order corrections $u_1^H=-q\omega\vartheta_H e_2^{(0)}/\mathcal D_L$ and $u_2^H=q\omega\vartheta_H e_1^{(0)}/\mathcal D_T$. With damping neglected, to first order in $\vartheta_H$ and to quadratic order in the curvature with the flat denominators~\eqref{eq:DLDT}, the time-averaged angular momentum density $\langle\ell_z\rangle=\rho\omega\,{\rm Im}(u_1^*u_2)$ is
\begin{equation}
\langle\ell_z\rangle=-\frac{\rho\vartheta_H\omega^2q^2|\eta|^2}{2\mathcal D_L\mathcal D_T}\left[P\left(\Delta K-\frac{q^2P}{\mathcal D_L}\right)-\frac{4\mu\rho\omega^2K_{12}^2}{\mathcal D_T}\right].
\label{eq:S_lz_general}
\end{equation}
In the principal frame, $K_{12}=0$,~\eqref{eq:S_lz_general} reduces to~\eqref{eq:Lz_general}, and in any other frame its first term is~\eqref{eq:Lz_general} with $\mathfrak K_1$ and $\mathfrak K_2$ replaced by $K_{11}$ and $K_{22}$. The last term is inertial because the transverse displacement in~\eqref{eq:S_zeroth_general} relaxes the off-diagonal geometric shear up to $e_2^{(0)}=-\rho\omega^2K_{12}\eta/\mathcal D_T$, so it drops from the ripple average at leading order in $\rho\omega^2/\mu q^2$. At finite damping the zeroth-order solution alone contributes $\langle\ell_z\rangle^{(0)}=\rho\omega q^2 PS|\eta|^2\,{\rm Im}\!\left[(\mathcal D_L^{*}\mathcal D_T)^{-1}\right]$, even in $\vartheta_H$ and hence in $\mathbf{B}$. It vanishes in any principal frame, averages to zero over isotropic ripples through the odd moment $\langle PS\rangle=0$, and is removed exactly by field antisymmetrization.

On a minimal saddle patch, $K_{11}=-K_{22}=\mathfrak{K}$ and $K_{12}=0$, the controlled leading-curvature result is
\begin{equation}
\langle\ell_z\rangle_{\rm saddle}=-\frac{2\rho\mu\vartheta_H\omega^2q^2|\eta|^2\mathfrak K^2[(\lambda+\mu)q^2-\rho\omega^2]}{\mathcal D_L^2\mathcal D_T}.
\label{eq:S_lz_saddle}
\end{equation}
For $\lambda\geq0$ it has the sign of $-\vartheta_H$ below the transverse resonance, provided $q^2\gg\mathfrak K^2$ and the curvature is approximately constant over a wavelength. The result holds on a patch and implies neither a global minimal-surface solution nor an infrared bound.

For a statistically isotropic shallow ripple field $h_0$ with correlation length $\xi$ and $q\xi\gg1$, the curvature $K_{ab}\simeq-\partial_a\partial_b h_0$ is constant over a phonon wavelength and~\eqref{eq:S_lz_general} holds patch by patch, a restriction removed in Sec.~V. Only second moments enter, and isotropy gives $\langle K_{11}^2\rangle=\langle K_{22}^2\rangle=3s$, $\langle K_{11}K_{22}\rangle=\langle K_{12}^2\rangle=s$, with vanishing odd moments, $s=\sigma_K^2/8$, and $\sigma_K^2=\langle K_{ab}K^{ab}\rangle$. To quadratic order in the curvature the Ricci shifts in the denominators are higher order and drop. Averaging \eqref{eq:S_lz_general} gives
\begin{equation}
\langle\!\langle\ell_z\rangle\!\rangle=\frac{2\rho\vartheta_H\omega^2q^2|\eta|^2 s\mathcal N}{\mathcal D_L^2\mathcal D_T^2},
\label{eq:S_lz_disorder}
\end{equation}
with $\mathcal N=2\lambda^2q^2 (\mu q^2-\rho\omega^2)+\lambda\mu q^2(3\mu q^2-2\rho\omega^2)+\mu(\mu^2q^4+2\mu q^2\rho\omega^2-2\rho^2\omega^4)$. \newline

With $x=\rho\omega^2/\mu q^2$ and $r=\lambda/\mu$, $\mathcal N/\mu^3q^4=2r^2(1-x)+r(3-2x)+1+2x-2x^2$ is a parabola in $r$ opening upward for $x<1$, with vertex at $r^*=-(3-2x)/[4(1-x)]<-\tfrac12$, and at $r=-\tfrac12$ it equals $x(5-4x)/2\geq0$. Hence $\mathcal N>0$ for all $r>-\tfrac12$, equivalently Poisson ratio $\nu=r/(r+2)>-\tfrac13$, throughout the sub-resonant window $0<\rho\omega^2<\mu q^2$, and the average~\eqref{eq:S_lz_disorder} is sign-definite. For $\rho\omega^2\ll\mu q^2$ it reduces to~\eqref{eq:lz_disorder}, whose sign is opposite to that of the saddle contribution~\eqref{eq:S_lz_saddle}. The isotropic moments are those of rank-one curvatures $K_{ab}\propto Q_aQ_b$ averaged over the direction of $\mathbf Q$. For $\lambda\geq0$, each such corrugation satisfies the stronger componentwise sign result of Sec.~V, while the isotropic local result above also covers $-\mu/2<\lambda<0$. A minimal saddle superposes two orthogonal cylinders of opposite curvature, and the cross term between them reverses the sign in~\eqref{eq:S_lz_saddle}. The average over a homogeneous ripple field removes this cross term.

In steady state, pumping balances the total in-plane damping $\Gamma_{\rm tot}=\Gamma_{\rm supp}+\Gamma_{\rm mag}+\cdots$ with an isotropic tangential damping channel  $j$ exerting $f_j^a=-\rho\Gamma_j\dot u^a$, and contributing 
\begin{equation}
\left.\partial_t\ell_\perp\right|_j=\rho\,[\mathbf u\times(-\Gamma_j\dot{\mathbf u})]\cdot\mathbf n=-\Gamma_j\ell_\perp.
\label{eq:S_loss_rate}
\end{equation}
The elastic and Hall stresses are symmetric and produce no internal torque density, so in steady state the drive and clamp provide the external reaction balancing the damping channels. The angular momentum of the tangential motion is $L_z=\int\dd A\,\langle\!\langle\ell_z\rangle\!\rangle(\mathbf n\cdot\hat{\mathbf z})$. Channel $j$ acts on material displaced by $\boldsymbol\xi=\mathbf u+\eta\mathbf n$ and exerts the torque $-\Gamma_j\int\dd A\,\rho\langle\boldsymbol\xi\times\dot{\mathbf u}\rangle\cdot\hat{\mathbf z}$. If the angular momentum it removes is transferred to it, it receives the opposite torque, which for the support is~\eqref{eq:torque}. A mode linewidth alone does not identify the fraction that each channel receives.

The tangential angular momentum $L_z$ differs from that of the full displacement $\boldsymbol\xi$, which adds $\rho(\mathbf u\times\dot\eta\mathbf n+\eta\mathbf n\times\dot{\mathbf u})$. These terms vanish along the local normal but contribute to laboratory $z$ at the same ripple order. Their time averages are equal, so~\eqref{eq:torque} contains half of their sum. For $\mathbf n\simeq(-\nabla h_0,1)$, that sum per projected area is
\begin{equation}
\delta\ell_z^H=-\rho\omega\operatorname{Im}\left[\eta^*(\partial_xh_0\,u_H^y-\partial_yh_0\,u_H^x)\right].
\label{eq:S_lever_arm}
\end{equation}
At the preferred clamped point (seed 17, 320 elements per side), $\delta L_z^H/L_z=-0.756$. Thus the support-transfer torque is $0.622\,\Gamma_{\rm supp}L_z$, while the full laboratory angular momentum is $0.244L_z$, both with the Hall sign. This calculation supplies the lever-arm correction, but does not evaluate boundary traction.
The linear strain \eqref{eq:strain_components} omits the von K\'arm\'an term $\tfrac12\partial_a\eta\,\partial_b\eta$ of the flat membrane. For a single harmonic drive this term contains a static part and a $2\omega$ part. Its direct interference with the linear curvature response averages to zero over a drive period, and its own angular-momentum contribution begins at fourth order in amplitude, while~\eqref{eq:S_lz_general} is quadratic. These power-counting facts do not test the nonlinear stability or the achievable drive amplitude.

\section{IV. Positive-curvature susceptibility}

To derive the spherical inverse propagators~\eqref{eq:EM_sphere}, we diagonalize the covariant operator~\eqref{eq:covariant_phonon_operator}. Tangent fields on a sphere of radius $R$ decompose into spheroidal and toroidal multipoles, $\nabla Y_{\ell m}$ and $\mathbf n\times\nabla Y_{\ell m}$, built from scalar spherical harmonics $Y_{\ell m}$. The Weitzenb\"ock identity $-\nabla^b\nabla_bv_a=(\Delta_{\rm H}v)_a-R_a{}^{b}v_b$, where $\Delta_{\rm H}$ is the Hodge Laplacian with eigenvalue $\ell(\ell+1)/R^2$ on both families and $R_a{}^b=\delta_a{}^b/R^2$, gives $-\nabla^b\nabla_b=[\ell(\ell+1)-1]/R^2$ on either family. With $\nabla^a\nabla_aY_{\ell m}=-\ell(\ell+1)Y_{\ell m}/R^2$, the operator~\eqref{eq:covariant_phonon_operator} with inertia and damping yields
\begin{equation}
R^2\mathcal D^{\rm sph}_{L}=(\lambda+2\mu)\ell(\ell+1)-2\mu-\rho R^2\omega^2-i\rho R^2\Gamma_L\omega,\qquad R^2\mathcal D^{\rm sph}_{T}=\mu(\ell-1)(\ell+2)-\rho R^2\omega^2-i\rho R^2\Gamma_T\omega,
\label{eq:S_sph}
\end{equation}
which at $\ell=2$ are~\eqref{eq:sphere_denominators}. A direct evaluation of the elastic energy for the axisymmetric fields of~\eqref{eq:sphere_ansatz}, with $P_2$ replaced by $P_\ell$, reproduces~\eqref{eq:S_sph} for $\ell\leq8$. At $q^2=\ell(\ell+1)/R^2$ the shift is $-2\mu K_G$ in both inverse propagators, with one $K_G$ from the Ricci term of~\eqref{eq:covariant_phonon_operator} and the other from the Laplacian acting on tangent vectors.
The $\ell=1$ modes are a toroidal rigid rotation and a spheroidal conformal dilatation with compression stiffness $2 (\lambda + \mu)/R^2$. \newline 

The Hall vertex carries the same geometric factor as the shear stiffness. For the axisymmetric multipoles of Sec.~II with $P_2$ replaced by $P_\ell$, the shear components are $e_1=U_\ell g_\ell/2R$ and $e_2=V_\ell g_\ell/2R$, with $g_\ell=\partial_\theta^2P_\ell-\cot\theta\,\partial_\theta P_\ell$, and
\begin{equation}
\int\dd\Omega\,g_\ell^2=(\ell-1)(\ell+2)\int\dd\Omega\,(\partial_\theta P_\ell)^2.
\label{eq:S_shear_content}
\end{equation}
Dividing~\eqref{eq:S_shear_content} by the displacement norm gives the shear factor $(q^2-2K_G)/4$, so the transverse stiffness is $\mu(q^2-2K_G)$ and the Hall coupling is $\vartheta_H\omega(q^2-2K_G)/2$. Their static ratio therefore keeps its flat value $\vartheta_H\omega/2\mu$ for every $\ell \ge 2$, and the curvature correction to the static Hall response enters only through the longitudinal inverse propagator $\mathcal D_L^{\rm sph}$. 

\section{V. Ripple scattering beyond the local approximation}

The local average of Sec.~III requires ripples longer than the drive wavelength. For shorter ripples the response must resolve the in-plane waves scattered by the ripples. Write $h_0(\mathbf r)=\sum_{\mathbf Q}h_{\mathbf Q}e^{i\mathbf Q\cdot\mathbf r}$ and take the drive $\eta=\eta_0e^{i(\mathbf q\cdot\mathbf r-\omega t)}$. At leading order in the slope $|\nabla h_0|$ the metric is flat and $K_{ab}=-\partial_a\partial_bh_0$, so the curvature strain $\eta K_{ab}$ has the Fourier components
\begin{equation}
s_{ab}(\mathbf p)=\eta_0h_{\mathbf Q}Q_aQ_b,\qquad \mathbf p=\mathbf q+\mathbf Q.
\label{eq:S_source}
\end{equation}
Each ripple component is a cylindrical corrugation that acts as a grating and transfers momentum $\mathbf Q$ to the drive. The in-plane equations are linear in $\eta$, so the angular momentum is fixed by the normal displacement, and the normal equation of~\eqref{eq:motion} fixes the force that produces it. In the frame $\hat{\mathbf n}=\mathbf p/p$, $\hat{\mathbf t}=\hat{\mathbf z}\times\hat{\mathbf n}$, the elastic stress and the Hall stress~\eqref{eq:hall_stress} give for the in-plane amplitudes at $\mathbf p$
\begin{equation}
\mathcal D_Lu_n+iH_pu_t=ipP_{\mathbf p}-\vartheta_H\omega p\,s_{nt},\qquad \mathcal D_Tu_t-iH_pu_n=2i\mu p\,s_{nt}+\tfrac12\vartheta_H\omega p(s_{nn}-s_{tt}),
\label{eq:S_2x2}
\end{equation}
with $P_{\mathbf p}=(\lambda+2\mu)s_{nn}+\lambda s_{tt}$, $H_p=\vartheta_H\omega p^2/2$, and $\mathcal D_{L,T}$ the flat denominators~\eqref{eq:DLDT} at wavevector $p$.

Cross terms between distinct ripple components vanish in a statistically homogeneous ensemble $\langle h_{\mathbf Q}h^*_{\mathbf Q'}\rangle\propto\delta_{\mathbf Q\mathbf Q'}$, or in a periodic spatial average under a single plane-wave drive, so the components contribute to $\ell_z$ without interference. A finite envelope can mix distinct incident wavevectors into the same outgoing mode, so this argument alone does not apply to a tapered clamped membrane. With $\varphi$ the angle between $\mathbf Q$ and $\mathbf p$ and $P_\varphi=(\lambda+2\mu)\cos^2\varphi+\lambda\sin^2\varphi$, the angular momentum to first order in $\vartheta_H$ and at zero damping is
\begin{equation}
\langle\!\langle\ell_z\rangle\!\rangle=-\frac{\rho\vartheta_H\omega^2|\eta_0|^2}{2}\sum_{\mathbf Q}\frac{Q^4\langle|h_{\mathbf Q}|^2\rangle p^2}{\mathcal D_L\mathcal D_T}\left[P_\varphi\cos2\varphi-\frac{p^2P_\varphi^2}{\mathcal D_L}-\frac{\mu\rho\omega^2\sin^22\varphi}{\mathcal D_T}\right],
\label{eq:S_lsum}
\end{equation}
the scattered counterpart of~\eqref{eq:S_lz_general}, to which each term reduces for $Q\to0$ with $K_{ab}=Q_aQ_bh_{\mathbf Q}$.
For every component,
\begin{equation}
P_\varphi\cos2\varphi-\frac{p^2P_\varphi^2}{\mathcal D_L}=-\frac{P_\varphi}{\mathcal D_L}\left[\rho\omega^2\cos^2\varphi+\big(2(\lambda+\mu)p^2-\rho\omega^2\big)\sin^2\varphi\right].
\label{eq:S_identity}
\end{equation}
For $\lambda\geq0$ and $\mu p^2>\rho\omega^2$, both zero-damping denominators are positive, and~\eqref{eq:S_identity} makes every nonzero term in~\eqref{eq:S_lsum} have the sign of $\vartheta_H$. \newline

\paragraph{Finite damping.}
The field-odd response at finite damping follows directly from~\eqref{eq:S_2x2}, retaining both products $u_n^{(0)*}u_t^H$ and $u_n^{H*}u_t^{(0)}$. Using $\alpha = \cos^2 \varphi$, $\beta=\sin^2\varphi$, $m=\rho\omega^2$, and $d_j=\rho\Gamma_j\omega$ as shorthands, denote the real parts of the inverse propagators by $A=(\lambda+2\mu)p^2-m$ and $B=\mu p^2-m$. The contribution of a ripple mode is
\begin{equation}
\ell^H_{\mathbf Q}=\rho\vartheta_H\omega^2|\eta_0|^2Q^4|h_{\mathbf Q}|^2\,\mathcal K_\Gamma(p,\varphi),\qquad \mathcal K_\Gamma=\frac{p^2\mathcal N_\Gamma}{2(A^2+d_L^2)(B^2+d_T^2)},
\label{eq:S_finite_damping_kernel}
\end{equation}
with the numerator
\begin{equation}
\mathcal N_\Gamma=P_\varphi\big\{\alpha(mB-d_Ld_T)+\beta\big([2(\lambda+\mu)p^2-m]B+d_Ld_T\big)\big\}+4\mu\alpha\beta(mA-d_Ld_T).
\label{eq:S_finite_damping_numerator}
\end{equation}
This result reduces to~\eqref{eq:S_lsum} when damping vanishes. For nonnegative $\Gamma_L,\Gamma_T$ and $\lambda\geq0$, every term of $\mathcal N_\Gamma$ is nonnegative if
\begin{equation}
c_T^2|\mathbf q+\mathbf Q|^2>\omega^2+\Gamma_L\Gamma_T.
\label{eq:S_subsonic}
\end{equation}
Indeed this inequality gives $mB>d_Ld_T$, $A\geq B>0$, and $2(\lambda+\mu)p^2-m>0$, so the total has the Hall sign whenever it is nonzero. The componentwise proof assumes neither isotropy nor Gaussian statistics and applies at leading height order to a single traveling drive under a periodic spatial average or in a homogeneous ensemble.

For a statistically isotropic height spectrum $S_h(Q)$,
\begin{equation}
\langle\!\langle\ell_z\rangle\!\rangle=\frac{\rho\vartheta_H\omega^2|\eta_0|^2\sigma_K^2}{\mu q^2}\,\frac{\int dQ\,Q^5S_h(Q)F(Q/q)}{\int dQ\,Q^5S_h(Q)},
\label{eq:S_iso}
\end{equation}
where $F=\mu q^2\langle\mathcal K_\Gamma\rangle_{\hat{\mathbf Q}}$ is the dimensionless kernel averaged over the direction of $\mathbf Q$ at fixed $Q$ and $\mathbf q$, reducing to the zero-damping expression~\eqref{eq:S_lsum} as $\Gamma_j\to0$, and $Q^5$ combines the two-dimensional measure with the four derivatives in $\sigma_K^2$. For $Q\ll q$,~\eqref{eq:S_iso} tends to the local average~\eqref{eq:S_lz_disorder}. For $Q\gg q$ and $\mu Q^2\gg\rho\omega^2$, $F\to[\lambda+\mu+\rho\omega^2/q^2](q/Q)^4/[2(\lambda+2\mu)]$, and
\begin{equation}
\langle\!\langle\ell_z\rangle\!\rangle\to\frac{(\lambda+\mu)q^2+\rho\omega^2}{2\mu(\lambda+2\mu)}\,\rho\vartheta_H\omega^2|\eta_0|^2\langle h_0^2\rangle.
\label{eq:S_short}
\end{equation}
At zero frequency and leading height order, $u_a=\eta_0\partial_a h_0$ cancels the strain of a spatially uniform normal displacement. At finite frequency this tangential motion has inertia and is not the unforced rigid-translation solution. An exact static vertical translation by $Z$ instead has $\eta=Z/\sqrt{1+|\nabla h_0|^2}$ and contravariant $u^a=Z\partial_a h_0/(1+|\nabla h_0|^2)$. The short-ripple limit above includes both the drive-gradient contribution and the inertial contribution. \newline

\begin{figure}[ht]
\centering
\includegraphics[width=0.94\textwidth]{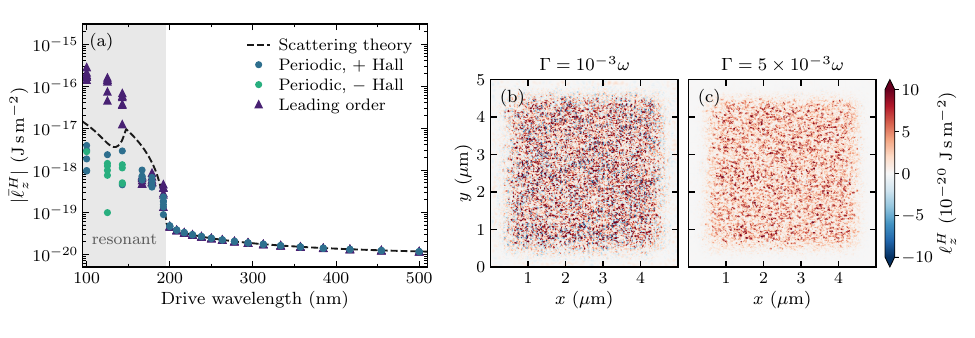}
\caption{\label{fig:S1}
(a) Magnitude of the mean field-odd angular momentum for $50$--$100\,\mathrm{nm}$ ripples at $10\,\mathrm{GHz}$ and $\eta_0=1\,\mathrm{nm}$. The line gives continuum theory~\eqref{eq:S_iso}, and circles show full-geometry periodic realizations, blue for the Hall sign and green for the opposite sign. Triangles show leading-height solutions on the same ripple modes, with opposite sign below $146\,\mathrm{nm}$. Shading marks primary scattering onto the transverse shell, where resonances invalidate prediction of individual realizations from the continuum average. (b),(c) Local density on the clamped $5\,\mu\mathrm m$ drum at the $250\,\mathrm{nm}$ drive (seed 17, 320 elements per side, $1\,\mu\mathrm m$ taper), for $\Gamma=10^{-3}\omega$ and $5\times10^{-3}\omega$. Drum-mode interference strongly changes the local pattern at lower damping but increases the mean only from $1.401$ to $1.431\times10^{-20}\,\mathrm{J\,s\,m^{-2}}$.}

\end{figure}

\paragraph{Drive spectrum and operating point.}
For a band $Q_1<Q<Q_2$ with $Q_1>q$,~\eqref{eq:S_subsonic} holds if $Q_1-q>\sqrt{\omega^2+\Gamma_L\Gamma_T}/c_T$. Table~\ref{tab:S_parameters} collects the parameters used below. A $100\,\mathrm{nm}$ drive scatters the $50$--$100\,\mathrm{nm}$ band across both acoustic shells. Full-geometry periodic solutions then have both angular-momentum signs among six realizations and peak linear strains of $7$--$9\%$ at $\eta_0=1\,\mathrm{nm}$, as shown in Fig.~\ref{fig:S1}(a). The leading-height approximation on the same modes can disagree even in sign because mean-square slope corrections of $0.009$ exceed the linewidth $\Gamma/\omega=0.001$ and shift resonances, so these points cannot support an off-resonant estimate. Lengthening the drive beyond $196\,\mathrm{nm}$ avoids the primary acoustic resonances for this band, and the main text uses $250\,\mathrm{nm}$. 

\begin{table}[ht]
\centering
\caption{Parameters shared by the periodic and clamped calculations, with the normal displacement imposed. The Hall coefficient uses the bulk-based five-layer scale of the main text, whose assumptions are discussed in Sec.~I.}
\label{tab:S_parameters}
\begin{ruledtabular}
\begin{tabular}{ll}
Quantity & Value\\
\hline
Lateral size $L$ & $5\,\mu\mathrm m$\\
Areal density $\rho$ & $1.2\times10^{-5}\,\mathrm{kg\,m^{-2}}$\\
Lam\'e coefficients $(\lambda,\mu)$ & $(20,50)\,\mathrm{N/m}$\\
Hall coefficient $\vartheta_H$ & $4\times10^{-14}\,\mathrm{kg\,s^{-1}}$\\
Drive frequency $\omega/2\pi$ & $10\,\mathrm{GHz}$\\
Drive amplitude $\eta_0$ & $1\,\mathrm{nm}$\\
In-plane damping $\Gamma_L=\Gamma_T$ & $10^{-3}\omega$\\
Ripple wavelengths & $50$--$100\,\mathrm{nm}$\\
Hessian-curvature rms $\sigma_K$ & $10^7\,\mathrm{m^{-1}}$\\
Preferred drive wavelength; taper width & $250\,\mathrm{nm}$; $1\,\mu\mathrm m$
\end{tabular}
\end{ruledtabular}
\end{table}

The reference height is a real random Fourier series with constant height power spectrum over the annulus $2\pi/(100\,\mathrm{nm})\leq Q\leq2\pi/(50\,\mathrm{nm})$. Independent complex Gaussian coefficients obey $h_{-\mathbf Q}=h_{\mathbf Q}^*$, with seeds 17--19 for the clamped calculations and 17--22 for the periodic ones. Each realization is normalized to $\langle(\partial_a\partial_bh_0)^2\rangle^{1/2}=10^7\,\mathrm{m^{-1}}$, which fixes the curvature variance at leading order in the slope, and the same Fourier coefficients are retained on every mesh. \newline

\paragraph{Numerical solution of the driven membrane response.}
We prescribe the normal displacement
\begin{equation}
\eta(\mathbf r,t)=\operatorname{Re}\left[\eta_0 e^{i(qx-\omega t)}w(x)w(y)\right],\qquad w(s)=t_s^3(10-15t_s+6t_s^2),\quad t_s=\min\{1,s/d,(L-s)/d\}.
\label{eq:S_taper}
\end{equation}
The polynomial is the quintic interpolant used in phase-field models of solidification~\cite{Wang1993}. It obeys $w=0$, $w'=w''=0$ at the edge and $w=1$, $w'=w''=0$ at the interior plateau, giving a $C^2$ envelope that enforces vanishing displacement and slope at the clamp without specifying an actuator model, and for $d=1\,\mu\mathrm m$ its projected mean square is $\langle[w(x)w(y)]^2\rangle=0.57261$. The finite envelope mixes incident wavevectors and permits interference, so the sign of the clamped response is tested by direct solution.

For $\mathbf X=(x,y,h_0)$, let $w_g=\sqrt{1+|\nabla h_0|^2}$, $g_{ij}=\delta_{ij}+\partial_i h_0\partial_j h_0$, and $K_{ij}=-\partial_i\partial_jh_0/w_g$. In terms of contravariant tangent components the exact linear strain is
\begin{equation}
E_{ij}=\tfrac12\left(g_{jk}\partial_i u^k+g_{ik}\partial_j u^k+u^k\partial_k g_{ij}\right)+\eta K_{ij}.
\label{eq:S_numerical_strain}
\end{equation}
In a local orthonormal frame write the engineering-strain vector as $\boldsymbol\varepsilon=(E_{\hat x\hat x},E_{\hat y\hat y},2E_{\hat x\hat y})^T=\mathsf B\mathbf u+\mathbf s$, where $\mathsf B$ maps tangential displacements to strain and $\mathbf s=\eta(K_{\hat x\hat x},K_{\hat y\hat y},2K_{\hat x\hat y})^T$ is the imposed curvature strain. The stiffness and Hall matrices acting on this vector are
\begin{equation}
\mathsf C=\begin{pmatrix}\lambda+2\mu&\lambda&0\\\lambda&\lambda+2\mu&0\\0&0&\mu\end{pmatrix},\qquad \mathsf V=\tfrac12\begin{pmatrix}0&0&-1\\0&0&1\\1&-1&0\end{pmatrix}.
\end{equation}
With the area-weighted adjoint $\mathsf B^\dagger$, the metric mass matrix $\mathsf M$, and equal damping rates $\Gamma=\Gamma_L=\Gamma_T$, the even and first-order Hall displacements solve
\begin{equation}
\mathsf D_0\mathbf u_0=-\mathsf B^\dagger\mathsf C\mathbf s,\qquad \mathsf D_0\mathbf u_H=i\omega\vartheta_H\mathsf B^\dagger\mathsf V(\mathsf B\mathbf u_0+\mathbf s),\qquad \mathsf D_0=\mathsf B^\dagger\mathsf C\mathsf B-\rho(\omega^2+i\Gamma\omega)\mathsf M.
\label{eq:S_numerical_system}
\end{equation}
The system~\eqref{eq:S_numerical_system} is exact in the ripple height, so it includes higher-order scattering of the induced in-plane waves off the static ripples. On the clamped drum, where both tangent components vanish at the edge, the system is solved with quadratic finite elements to a relative residual below $2\times10^{-11}$.
The observable is the tangent-motion contribution projected on laboratory $z$, per projected area,
\begin{equation}
\bar\ell_z^H=\frac{\rho\omega}{L^2}\int dx\,dy\,w_g\,\operatorname{Im}\left[(u_0^x)^*u_H^y+(u_H^x)^*u_0^y\right].
\label{eq:S_numerical_observable}
\end{equation}
The use of contravariant Monge components makes $(\mathbf u\times\dot{\mathbf u})\cdot\hat{\mathbf z}=u^x\dot u^y-u^y\dot u^x$. 
On a periodic square of the same side, the system~\eqref{eq:S_numerical_system} is solved spectrally. Derivatives are evaluated by fast Fourier transform on a $512^2$ grid, and GMRES iteration about the flat membrane converges to a relative residual of order $10^{-9}$. 
Doubling the grid changes $\bar\ell_z^H$ by less than one part in $10^5$. Drive wavelengths are $L/n$ with integer $n$, a uniform plane-wave drive isolates the bulk response, and the taper~\eqref{eq:S_taper} reproduces the clamped forcing without the clamp. 

\begin{table}[ht]
\centering
\begin{ruledtabular}
\begin{tabular}{rrrr}
Seed & Elements per side & $\bar\ell_z^H$ at $250\,\mathrm{nm}$ & $\bar\ell_z^H$ at $312.5\,\mathrm{nm}$\\
\hline
17 & 256 & 1.54197 & 1.07574\\
17 & 320 & 1.43113 & 1.03718\\
17 & 384 & 1.43512 & 1.03969\\
18 & 256 & 1.49681 & 1.12066\\
18 & 320 & 1.42519 & 1.03162\\
18 & 384 & 1.43111 & 1.03355\\
19 & 256 & 1.49324 & 1.07546\\
19 & 320 & 1.40911 & 1.01269\\
19 & 384 & 1.41153 & 1.01735
\end{tabular}
\end{ruledtabular}
\caption{Direct clamped-membrane results at $\eta_0=1\,\mathrm{nm}$, $d=1\,\mu\mathrm m$, and the material parameters given in the text. Densities are in $10^{-20}\,\mathrm{J\,s\,m^{-2}}$, with the uncorrected mesh values retained.}
\label{tab:S_clamped}
\end{table}

\paragraph{Convergence and comparison with scattering theory.}
Periodic full-geometry solutions confirm the scattering theory, averaged over ripple realizations, $\bar\ell_z^H$ agrees with~\eqref{eq:S_iso} within $2\%$ from $208$ to $500\,\mathrm{nm}$ and deviates only near the $196\,\mathrm{nm}$ threshold, where scattered waves approach the transverse resonance, see Fig.~\ref{fig:numerics}(c). Because $\bar\ell_z^H$ sums $\sim10^4$ independent ripple contributions, its standard deviation between realizations is about $1\%$ of the mean, and a single membrane gives the ensemble value. The exact geometry changes $\bar\ell_z^H$ from its leading-height value by at most $1.6\%$, of order the mean-square slope, so the leading-order theory is controlled.
\begin{table}[ht]
\centering
\caption{Clamped results at 384 elements per side and comparisons with the bulk continuum prediction. Densities are in $10^{-20}\,\mathrm{J\,s\,m^{-2}}$, and the envelope factor is $\langle[w(x)w(y)]^2\rangle=0.57261$. Sample spreads are standard deviations divided by the mean; mesh changes are the largest over the three seeds.}
\label{tab:S_comparison}
\begin{ruledtabular}
\begin{tabular}{lrr}
Quantity & $250\,\mathrm{nm}$ & $312.5\,\mathrm{nm}$\\
\hline
Clamped mean density & 1.426 & 1.030\\
Sample spread & $0.9\%$ & $1.1\%$\\
Change from 320 to 384 elements per side & $0.42\%$ & $0.46\%$\\
Continuum density, uniform drive & 2.404 & 1.740\\
Clamped / (continuum $\times$ envelope) & 1.036 & 1.034\\
Tapered periodic / (continuum $\times$ envelope) & 1.011 & 1.001\\
Clamped / tapered periodic & 1.025 & 1.033
\end{tabular}
\end{ruledtabular}
\end{table}
Clamped drums, which include the finite drive envelope and fixed edges of an experiment, confirm the prediction at the $250\,\mathrm{nm}$ operating point: their mean $\bar\ell_z^H$ has the Hall sign and exceeds it at equal mean-square normal displacement by $3.6\%$ (Table~\ref{tab:S_comparison}). The excess comes mainly from resonant in-plane drum modes, which carry the Hall sign in every computed case and are excited by the envelope's spectral tails and by higher-order scattering. A fivefold larger damping suppresses these modes and brings the clamped $\bar\ell_z^H$ within $0.03\%$ of its tapered-periodic value as shown in Fig.~\ref{fig:S1}(c). Peak linear strains are $0.55$--$0.67\%$, and the last mesh refinement changes $\bar\ell_z^H$ by less than $0.5\%$.\newline

\paragraph{Torque scale and amplitude dependence.}

The preferred result gives $L_z=L^2\bar\ell_z^H=3.56\times10^{-31}\,\mathrm{J\,s}\simeq3.4\times10^3\hbar$. At $\Gamma_{\rm supp}=10^{-3}\omega$, the tangential transfer scale is $\Gamma_{\rm supp}L_z=2.24\times10^{-23}\,\mathrm{N\,m}$; the lever arm of Sec.~III reduces the conditional torque~\eqref{eq:torque} to $1.39\times10^{-23}\,\mathrm{N\,m}$. Doubling the amplitude gives four times the torque, $5.57\times10^{-23}\,\mathrm{N\,m}$, and a largest linear strain of $1.33\%$. 
These rescalings use linear displacement theory and do not include nonlinear elasticity or the actuator force.

